# Long-Term Memory: Scaling of Information to Brain Size

Donald R. Forsdyke

**Abstract**  The material bases of information – paper, computer discs – usually scale with information quantity. Large quantities of information usually require large material bases. Conventional wisdom has it that human long-term memory locates within brain tissue, and so might be expected to scale with brain size which, in turn, depends on cranial capacity. Large memories, as in savants, should always require large heads. Small heads should always scale with small memories. While it was previously concluded that neither of these predictions was invariably true, the evidence was weak. Brain size also depends on ventricle size, which can remain large in some survivors of childhood hydrocephaly, occupying 95% of cranial volume. Yet some of these have normal or advanced intelligence, indicating little impairment of long-term memory. This paradox challenges the scaling hypothesis. Perhaps we should be looking further afield?

**Keywords** Computer metaphor . Connectionist paradigm . Head size. Information capacity . Plasticity limits . Ventricle size

D. R. Forsdyke

Department of Biomedical and Molecular Sciences,

Queen's University, Kington, Ontario, Canada K7L3N6

e-mail: forsdyke@queensu.ca

## Introduction

> About the truth and extent of these facts none but men possessing a special knowledge of
> physiology and natural history have any right to an opinion; but the superstructure based on
> those facts enters the realm of pure reason, and may be discussed apart from all doubt as the
> fundamental facts (Jenkin 1867, p. 278).

With little modification, these remarks on Darwin's evolution theory by of a professor of
engineering with little knowledge of biology, will serve to justify the present foray of one not
possessing a special knowledge of neurophysiology into the domain of memory.

## Size Does Not Scale with Information Content

Mother Hubbard spoke authoritatively when she declared the cupboard bare. She knew the size
and form of bones likely to satisfy her dog. She also knew the size of her cupboard. Furthermore,
she knew that bones were stable, uncompressible and, most importantly, likely to be needed at
short notice by her canine friend. We cannot speak on human long-term memory with equal
authority. It often appears stable and can be called upon at short notice, but beyond these facts
we cannot now go. We can, however, speak about the size of the metaphorical 'cupboard' where,
conventional wisdom holds, long-term memory must lie (Draaisma 2000). And if the cupboard
proves to be bare, we have to admit a problem (Forsdyke 2009).

The cranial 'cupboard' might be expected to be larger than normal in individuals (savants)
who seem to have very large long-term memories. Although some savants are deficient in other
respects, the discovery of just one savant without serious deficiency and with normal cranial
capacity would refute this prediction. It so happens that, although one much celebrated savant
had a large head, most savants have not, and some show no serious deficiency (Treffert 2010).
Conversely, when the cranium is much smaller than normal (microcephaly), long-term memory
might be decreased. Unfortunately, microcephalics tend to be tested more for intelligence
(usually impaired), than for their ability to recall information. However, on the assumption that
intelligence and long-term memory are to some extent related, we can note that a few
microcephalics have normal intelligence (Forsdyke 2009). Thus, with both savants and microcephalics, the evidence, albeit weak, suggests a *disconnect* between the volume of neural tissue held within the cranium and the quantity of information which that neural tissue is, in some way, held to store. *Size does not matter*. Studies of hydrocephalics are casting new light on this scaling paradox.

**Adult Hydrocephalics with 5% Brain Tissue by Volume**

With savants and microcephalics, the volume of neural tissue is determined by cranial capacity. However, with hydrocephalics the volume is largely determined by the size of the fluid-filled ventricles. A therapeutic shunt in early life can lower a cerebrospinal fluid pressure that otherwise would relentlessly compress neural tissue against the cranial surface. In the 1970s innovations in non-invasive brain-scanning technology facilitated the reexamination in adult life of treated hydrocephalics. The journal *Science*, under the title "Is your brain really necessary?" (Lewin 1980), described a series of 600 cases with residual ventricular enlargement that had been studied in Britain by paediatrician John Lorber (1915-1996). Again, while long-term memories were not directly assessed, intelligence quotients (IQs) were.

   Amazingly, in 60 of Lorber's cases, ventricular fluid still occupied 95% of cranial capacity. Yet half of this group had IQs above average. Among these was a student with an IQ of 126 who had a first class honours degree in mathematics and was socially normal. For this case Lorber noted:

> Instead of the normal 4.5 centimetre thickness of brain tissue between the ventricles and the cortical surface, there was just a thin layer of mantle measuring a millimeter or so. The cranium is filled mainly with cerebrospinal fluid. … I can't say whether the mathematics student has a brain weighing 50 grams or 150 grams, but it's clear that it is nowhere near the normal 1.5 kilograms.

   Lorber's findings met much scepticism (Lewin 1980). But recently there have been two independent confirmations, suggesting Lorber should not have been so lightly dismissed. Under the title, "Brain of a white-collar worker," French neurologists (Feuillet et al. 2007) showed "massive ventricular enlargement" in the brain scan of a civil servant who had an IQ in the low normal range and came to them with relatively mild neurological symptoms that responded to treatment. Shortly thereafter, neurosurgeons in Brazil reported a similar case (Oliviera et al. 2012). Their figure (Fig. 1) is striking, since it compares the enlarged ventricles of their symptom-free subject with the equally enlarged ventricles of a subject with "deep cognitive and motor impairment" – the more usual expectation.

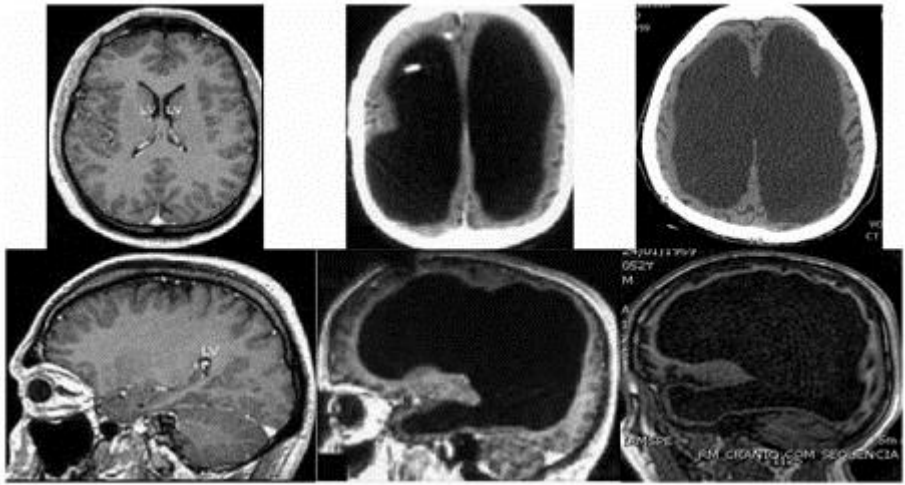

**Fig. 1.** Brain scans of patients of Oliveira et al. (2012). Normal adult appearance (left), with "LV" referring to the small black fluid-filled ventricles. Enlarged ventricles (middle and right). The middle patient is clinically normal, whereas the right patient has had "deep cognitive and motor impairment since childhood." Reproduced with author's permission from *Frontiers in Human Neuroscience* under Creative Commons Licence.

## Plasticity or Paradox?

Data from savants, microcephalics and hydrocephalics seem to be telling us that, with respect to long-term memory, there are circumstances where, paradoxically, size does not matter. Lorber suggested that "primitive deep structures that are relatively spared in hydrocephalus" may have allowed his subjects to live normal lives, so that "there must be a tremendous amount of redundancy or spare capacity in the brain." This implied that normally much of the brain is simply idling, ready to act as a backup should the need arise.

Somewhat more convincing was a "plasticity" explanation advanced by Bateson and Gluckman (2011, p. 32-34) when commenting on the French patient. In similar fashion, the Brazilian team invoked the "resilient adaptation of brain networks" associated with "the ability of neuronal tissue to reassume and reorganize its functions" (Oliviera et al. 2012). These plasticity explanations imply that, in keeping with the sometimes amazing recoveries reported for severe brain injuries, an otherwise-occupied part of the brain can change to compensate for a defective part. Under the prevailing neural network "connectionist" paradigm (Draaisma 2000; Forsdyke 2009), for hydrocephalics with only 5% of neural tissue remaining, this would seem to require the establishment early in life of a critical network, which would have to retain its connectivity in the face of the subsequent severe progressive distortion associated with ventricular expansion. Unfortunately, there are no post-mortem histological studies on this.

However, there must be rules for redundancy and plasticity. There *must be limits*. It is a matter of elementary logic that, at some stage of brain shrinkage, these explanations must fail (Majorek 2012). The drastic reduction in brain mass in certain, clinically-normal, hydrocephalic cases, seems to demand unimaginable levels of redundancy and/or plasticity – *superplasticity*. How much brain must be absent before we abandon these explanations and look elsewhere? Perhaps we are looking a gift paradox in the mouth?

The extent of our neurobiological ignorance was recently noted by Eric Kandel (2006, p. 423): "In the study of memory storage, we are now at the foothills of a great mountain range. … To cross the threshold from where we are to where we want to be, major conceptual shifts must take place." Regarding the human brain's "massive storage capacity" for object details, Brady et al. (2008) have also challenged "neural models of memory storage and retrieval." Others are calling for "radical modification of the standard model of memory storage" (Fusi and Abbott 2007; Firestein 2012). Given the doubts of these specialists, perhaps it is time for other hypotheses to be admitted to the table of responsible neuroscientific discourse. Should we not be looking further afield – alert for rare gift horses (Heerden 1968; Pribram 1991; Talbot 1991; Berkovich 1993)? Metaphors may help.

**From Stand-Alone to Cloud Computing**

In 1934 librarian Paul Otlet (1868-1944) envisioned a "mechanical collective brain" to which individuals could connect through "electric telescopes" that would seem to equate with today's personal computers (Levie 2006). And in 1970 at the first international conference on "Man and Computer," computer engineer John McCarthy (1972) sketched out how the "home informational terminal" – a console – might one day retrieve personal files from, and store personal files to, a central resource – now known as "the cloud." However, when personal computers appeared in the 1980s they were essentially stand-alone, with their own software and data-storage (memory). It was easy to relate this, metaphorically, to the perception that an individual human brain is a stand-alone entity, with its own software and memory (Draaisma 2000; Noll 2003). Indeed, concerning "generalized cognition-space," Lenneberg (1965, pp. 257-261) had pointed to doubts raised by logical analogy:

> There is, however, another line of argument that induces many scholars to suspect a close relationship between brain size and intelligence. It is based on purely logical considerations; in fact, the reasoning underlying it is by analogy. The capacities of an electronic computer or desk calculator are directly related to the number of its constituent elements. This engenders the belief that an increase in the number of units in the brain has a similar consequence. However, evidence on this is surprisingly poor. … Although it is entirely possible that the emergence of language and intelligence are historically related to the increase in size of the brain, the case is certainly not yet irrefutably proven.

The known stand-alone memory storage alternatives have been reviewed elsewhere. For example, although speculation continues, the idea that long-term memory might reside in brain DNA is largely put to rest (Forsdyke 2009). The question of a role for non-DNA polymers remains. Broadly, this category includes other macromolecules (RNA, protein, lipid, carbohydrate), or unknown subatomic forms perhaps related to "quantum computing" (Sciarrino and Mataloni 2012). Also in this category is the attractive idea of the brain as a three dimensional holographic storage network (Heerden 1968; Pribram 1991; Draaisma 2000). But, before seeking such exotic storage modalities, we should ensure that the brain "cupboard" is indeed bare of forms more in keeping with current paradigms. We need a better inventory of brain-specific macromolecules to exclude a polymeric form that, DNA-like, might store information digitally (Tsien 2013). Furthermore, although brain information might be stored in some subatomic form, at some point that form would need to interface with more conventional macromolecular species (e.g. proteins). Specific adaptations for this role should distinguish them from other macromolecules.

For all these storage alternatives, the thinking is conventional in that long-term memory is held to be *within* the brain, and the hydrocephalic cases remain hard to explain. Yet currently most of us, including the present author, would prudently bet on one or more of the stand-alone forms. The unconventional alternatives are that the repository is external to the nervous system, either elsewhere within the body, or extra-corporeal. The former is unlikely since the functions of other body organs are well understood. Remarkably, the latter has been on the table since at least the time of Avicenna and hypothetical mechanisms have been advanced (Talbot 1991; Berkovich 1993; Forsdyke 2009; Doerfler 2010). Its modern metaphor is "cloud computing."

Even though the internet emerged in the 1990s (Berners-Lee 2010), it took two decades for cloud computing to become established (Furht 2010). Imaginative attempts to relate this to the workings of individual brains (Talbot 1991; Berkovich 1993), still fall far short on evidence (Forsdyke 2009). However, the rare hydrocephalic cases described here suggest we should exercise caution when tempted to cast aside the astonishing idea of personal information – long-term memory – being remotely stored. After all, Nature is not obliged to conform to our preconceptions. And, as Sherlock Holmes once said, *"*when you have eliminated the impossible, whatever remains, however improbable, must be the truth."

The importance of this extends far beyond neuroscience and the clinic. When speaking of extracorporeal memory storage we enter the domain of "mind" or "spirit," with corresponding metaphysical implications (Crick 1995; Draaisma 2000; Forsdyke 2009). We begin to "secularize the soul" (Hacking 1995). Perhaps we should return to 1867 and harken to an exchange between Robert Chambers and Alfred Russel Wallace (Wallace 1905, p. 286): "The term 'supernatural' is a gross mistake. We have only to enlarge our conceptions of the natural, and all will be alright."

Thus, there may be vestiges of truth amongst the dross that we poor creatures, imprisoned within the second decade of the twenty-first century, can comprehend no better than those imprisoned in the later decades of the nineteenth century would have comprehended Gregor Mendel, had they known of him (Cock and Forsdyke 2008, p. 197-264). And that which is now deemed metaphorical may not always remain so. Draaisma (2000) notes that metaphors can die and become literal. There are those who urge us to lift our eyes to new horizons (Talbot 1991; Berkovich 1993). While they may lack a formal training in neuroscience, we should listen carefully.

**Acknowledgements**  Timothy Crow, David Murray and Hans Noll kindly reviewed the text. Queen's University hosts my evolution education webpages where some cited articles may be found (http://post.queensu.ca/~forsdyke/mind01.htm ). YouTube hosts a lecture based on this paper: http://www.youtube.com/watch?v=fA9Sdzi4ZOU

**References**

Bateson P, Gluckman P (2011) Plasticity, robustness, development and evolution. Cambridge University Press, Cambridge

Berkovich SY (1993) On the information processing capabilities of the brain: shifting the paradigm. Nanobiol 2:99–107

Berners-Lee T (2010) Long live the web. Sci Amer 303:80–85

Brady TF, Konkle T, Alvarez GA, Oliva A. (2008) Visual long-term memory has a massive storage capacity for object details. Proc Natl Acad Sci USA 105:14325–14329

Cock AG, Forsdyke DR (2008) Treasure your exceptions. The science and life of William Bateson. Springer, New York

Crick F (1995) The astonishing hypothesis. The scientific search for the soul. Touchstone Books, London

Doerfler W (2010) DNA – a molecule in search of additional functions: recipient of pool wave emissions? A hypothesis. Med Hypoth 75: 291-293

Draaisma D (2000) Metaphors of memory. A history of ideas about the mind. Cambridge University Press, Cambridge

Feuillet L, Dufour H, Pelletier J (2007) Brain of a white-collar worker. Lancet 370:362

Firestein S (2012) Ignorance. How it drives science. Oxford University Press, New York

Forsdyke DR (2009) Samuel Butler and human long-term memory. Is the cupboard bare? J Theor Biol 258:156–164

Furht B (2010) Cloud computing fundamentals. In: Furht B, Escalente A (eds) Handbook of cloud computing. Springer, New York, pp 3–19

Fusi S, Abbott LF (2007) Limits on the memory storage capacity of bounded synapses. Nature Neurosci 10:485–92

Hacking I (1995) Rewriting the soul. Multiple personality and the sciences of memory. Princeton University Press, Princeton

Heerden PJ van (1968) The foundation of empirical knowledge. N. V. Uitgeverij Wistik, Wassenaar, p. 46

Jenkin F (1867) The origin of species. North Brit Rev 46:277–318

Kandel ER (2006) In search of memory. Norton, New York

Lenneberg EH (1965) Biological foundations of language. Wiley, New York

Levie F (2006) L'Homme qui voulait classer le monde: Paul Otlet et le Mundaneum. Les Impressions Nouvelles, Paris-Bruxelles

Lewin R (1980) Is your brain really necessary? Science 210:1232–1234

Majorek MB (2012) Does the brain cause conscious experience? J Consc Stud 19: 121-144

McCarthy J (1972) The home information terminal. In: Marois M (ed)  Man and computer. Proceedings of international conference, Bordeaux 1970. Karger, Basel, pp. 48–57

Noll H (2003) The digital origin of human language – a synthesis. BioEssays 25:489–500

Oliviera MF de, Pinto FCG, Nishikuni K, Botelho RV, Lima AM, Rotta JM (2012) Revisiting hydrocephalus as a model to study brain resilience. Front Hum Neurosci 5: article 181, 1-4

Pribram KH (1991) Brain and perception. Lawrence Erlbaum, Hillsdale, pp. xxii–xxiv, pp. 277–278

Talbot M (1991) The holographic universe. Harper Collins, New York

Treffert DA (2010) Islands of Genius. Jessica Kingsley, London

Tsien RY (2013) Very long-term memories may be stored in the pattern of holes in the perineuronal net. Proc Natl Acad Sci USA 110: 12456–12461

Sciarrino F, Mataloni P (2012) Insight on future quantum networks. Proc Natl Acad Sci USA 109:20169–20170

Wallace AR (1905) My life. A record of events and opinions, Volume 2. Chapman and Hall, London